\documentstyle[floats,twocolumn,aps,prl]{revtex}

\tighten

\begin{document}
\draft
\title{
\vspace*{-1cm}\hfill
 {\tt }
       \vspace{1cm}\\
Adlayer core-level shifts of random metal overlayers on
transition-metal substrates}

\author{M.V. Ganduglia-Pirovano$^{1,2,3}$, 
J. Kudrnovsk\'{y}$^{4}$ and M. Scheffler$^1$}
\address{$^1$Fritz-Haber-Institut der Max-Planck-Gesellschaft, Faradayweg
4-6, D-14$\,$195 Berlin-Dahlem, Germany}
\address{$^2$Max Planck Institut f\"ur Physik komplexer Systeme,
Au\ss enstelle Stuttgart,
Heisenbergstrasse 1, D-70$\,$569 Stuttgart, Germany}
\address{$^3$CAMP 
Department of Physics, Technical University of Denmark, 
DK-2800 Lyngby, Denmark}
\address{$^4$
Institute of Physics, Academy of Sciences of the Czech Republic,
CZ-180$\,$40 Prague 8, Czech Republic}

\date{\today}

\twocolumn[
\maketitle

\vspace*{-10pt}
\vspace*{-0.7cm}
\begin{quote}
\parbox{16cm}{\small


We calculate the difference of the
ionization energies of a core-electron of a surface alloy,
i.e., a B-atom in a A$_{1-x}$B$_{x}$ overlayer on a fcc-B(001)-substrate,
and a
core-electron of the clean fcc-B(001) surface
using density-functional-theory.
We analyze the initial-state contributions and the screening
effects induced by the core hole, and
study
the influence of the alloy composition for a number of noble metal-transition
metal systems.
Data are presented for Cu$_{1-x}$Pd$_{x}$/Pd(001),
Ag$_{1-x}$Pd$_{x}$/Pd(001),
Pd$_{1-x}$Cu$_{x}$/Cu(001), and Pd$_{1-x}$Ag$_{x}$/Ag(001), changing
$x$ from 0 to 100 \%.
Our analysis clearly indicates the importance of final-state
screening effects for the interpretation of measured 
core-level shifts. Calculated deviations from the
initial-state trends are explained in terms of the change of inter- 
and intra-atomic screening upon alloying.
A possible role of alloying on the chemical reactivity of metal surfaces
is discussed.

\pacs{PACS numbers: 79.60.Dp 73.20.At 73.61.At}
}
\end{quote} ]

\narrowtext

Surfaces of metallic alloys can act as efficient catalysts \cite{sinfelt}.
The yield of
such chemical reactions depends sensitively on the structure and the
composition of these surfaces. 
Recently, 
in trying to understand the role 
of the electronic interactions between the components of bimetallic 
catalysts,
the shift of core-electron binding energies of surface atoms
of metal adlayers on transition-metal substrates have been studied, both
experimentally \cite{goodman1} and theoretically
\cite{weinert,wu,henn96}. It has been demonstrated 
that when final-state effects, i.e., contributions due to differences in
final-state relaxation energies, do not modify the initial-state trend,   
core-level shifts
can be used to deduce information on the adlayer's
$d$ density of states (DOS) \cite{weinert,wu,henn96}.
Indeed, for the studied adlayer systems,
the shift of core-electron eigenvalues sensitively detected the
shift of the valence $d$ band at the surface. The latter is of particular
importance because the position of the surface $d$ band relative to the
Fermi level is a key factor determining the chemical
reactivity of the surface \cite{hamm95a,hammer95}.

In this Letter we extend these studies to 
random metallic overlayers on  non-random 
metallic substrates (so-called surface alloys).
In particular, we investigate the trends in variations of the core-level 
shifts of the surface atoms as a function of concentration of foreign
atoms in the surface layer.
We focus our attention on systems formed by the combination of transition
and noble metals because of their different capabilities to screen
the core-hole created in the core-level photoemission spectroscopy.
The study of the concentration dependence of the
different contributions to the shifts allows for a deeper understanding of
the physical mechanism  underlying  final-state effects on the core-level
shifts at metal surfaces.
We  show that  the trends in variations in
these shifts are not always given by the initial-state shifts as in
studies of non-alloyed surfaces 
\cite{henn96,skriver94,meth95,pehl93}. We have identified two 
examples for which changes in the screening of the
core-hole due to modifications of the chemical environment of
the adatoms result in qualitatively different trends.

We present a density-functional-theory study 
of the trends  of the initial-state and final-state screening contributions
to the adlayer core-level shifts for surface alloys.
We define
the adlayer core-level shift (ACLS) as the difference    
of the ionization energies of an adlayer
core-electron of a B-atom in a random
overlayer A$_{1-x}$B$_{x}$ on a B-substrate
and of a core-electron
of the clean surface of the B-metal.
Thus, using the system
Cu$_{1-x}$Pd$_{x}$/Pd(001) as an example, the ACLSs  of a Pd atom are 

\begin{eqnarray}
\Delta^\prime_{\rm ACLS} & = &
\tilde{I}_{c}^{{\rm Cu}_{1-x}{\rm Pd}_{x}/{\rm Pd(001)}}
-\tilde{I}_{c}^{\rm Pd(001)}\nonumber\\
&=& E^{{\rm Cu}_{1-x}{\rm Pd}_{x}/{\rm Pd(001)}}(n_{c}=1)\nonumber\\
& &-E^{{\rm Cu}_{1-x}{\rm Pd}_{x}/{\rm Pd(001)}}(n_{c}=2) \nonumber \\
& &-E^{\rm Pd(001)}(n_{c}=1)
+E^{\rm Pd(001)}(n_{c}=2) \quad , \label{eq1}
\end{eqnarray}
\noindent where $\tilde{I}_c$ are the core-level ionization energies
with respect to the Fermi energy,
and $E$ are
total energies of the ground state (two electrons in the core level,
$n_{c}=2$) and the excited state
($n_{c}=1$),  calculated under the constraint of overall charge neutrality. 
The upper indices identify the considered systems, namely the adsorbed 
Cu$_{1-x}$Pd$_{x}$ random monolayer on a Pd(001) substrate and the 
top layer of the clean Pd(001) surface.
The  prime in $\Delta^\prime_{\rm ACLS}$ is to note that our definition here
slightly differs from that introduced in Ref. \cite{henn96}.
In Ref. \cite{henn96}
admetal monolayers of A-atoms on a B-substrate were considered
and the ionization energies of the A-atoms were compared with those of a
clean surface of the A-crystal. 
Here, we calculate the difference of the
ionization energies of a core-electron of 
B-atoms in a A$_{1-x}$B$_{x}$ overlayer on a B-substrate
and a
core-electron of a clean surface of the B-crystal. Thus, 
the full monolayer situation of A on B ($x\rightarrow 0$) 
means the limit of zero concentration of B atoms at random 
overlayer.

Using the Slater-Janak transition-state concept
to evaluate total-energy
differences \cite{ts}, we obtain from Eq. (\ref{eq1})
\begin{eqnarray}
\Delta_{\rm ACLS}^\prime & \approx &
-\epsilon_{c}^{{\rm Cu}_{1-x}{\rm Pd}_{x}/{\rm Pd(001)}}(n_{c}=1.5) \nonumber\\
& & +\epsilon_{c}^{\rm Pd(001)}(n_{c}=1.5) \quad , \label{eq2}
\end{eqnarray}
\noindent where
$\epsilon_{c}^{{\rm Cu}_{1-x}{\rm Pd}_{x}/{\rm Pd(001)}}$
 and $\epsilon_{c}^{\rm Pd(001)}$
denote the
Kohn-Sham eigenvalues of a particular core state of an adlayer Pd atom
and
a surface atom of clean Pd(001).
In the initial-state approximation, the ACLSs
are given by Eq. (\ref{eq2}) with $n_c=2$.

The calculations of the ACLSs are performed by means of 
the surface Green's function technique based on the all-electron 
tight-binding linear muffin-tin orbital method (LMTO) within the 
atomic-sphere and the local-density approximations
(LDA)
combined
with the single-site coherent potential approximation (CPA) in order to treat
the effect of disorder \cite{josef1}. Recently, the method 
has been successfully 
applied to the study of
surface core-level shifts (SCLSs)
of the $4d$ transition metal surfaces \cite{clean}.
A related approach 
has been applied to the study of the  SCLSs
of the $4d$ and $5d$ transition metal surfaces \cite{skriver94}. 
Though they have included final-state effects
because of the use of the
frozen-core approximation, they are not able to separate initial and 
final-state contributions to the shifts.
The present approach is ideally suited to the nature of the problem since the
whole concentration range, from the monolayer to the single
impurity limit, can be treated on an equal footing
as can the neutral system and the system with a core-hole
at a single surface atom.
The potentials are calculated self-consistently
with respect to both CPA and LDA
in a
region
consisting of the overlayer or the
surface layer, three substrate layers, and two layers of
empty spheres 
at the vacuum-sample interface. This 
region is coupled via the surface Green's
function technique to the semi-infinite vacuum on one side
and to the semi-infinite crystal on the other, with frozen potentials.
An ideal epitaxial growth has been assumed, that means that all
interatomic distances in the overlayer, as well as between overlayer and
substrate are assumed to be the same and equal to that in the substrate.
Calculations are performed for sphere radii chosen so as to minimize
the total energy of the bulk substrate in the fcc structure.
To describe the transition-state (see Eq. \ref{eq2}), separate
LDA self-consistent
calculations of a single-impurity, with half an 
electron missing in the particular core-level, at the random
overlayer and at the clean surface
are performed under the constraint of overall charge neutrality.

For the random overlayer case, 
the properties of the
individual atoms occupying the ideal lattice sites are characterized by
the coherent potential function matrix for the corresponding
non-ionized system.
Strictly speaking, this is an approximation as we would have to perform the
configurational averaging with the core-ionized atom  fixed at a given
lattice site at the random overlayer which would result in an 
inhomogeneous CPA medium centered around the impurity. 
Thus, the use of the calculated CPA
medium that corresponds to the non-ionized system 
with the same overlayer composition, is the simplest meaningful
approximation, which averages over the different configurations.

Figure 1(a) shows the calculated ACLS for 
Cu $2p$ electrons in the random 
Pd$_{1-x}$Cu$_{x}$ overlayer on Cu(001) and Fig. 1(b) 
for Pd $3d$ electrons  in
Cu$_{1-x}$Pd$_{x}$ on Pd(001). The results for 
Ag $3d$ electrons in the random   Pd$_{1-x}$Ag$_{x}$ overlayer on 
Ag(001) and 
for Pd $3d$ electrons 
in  Ag$_{1-x}$Pd$_{x}$ on Pd(001) are qualitative very similar
to those of Figs. 1(a) and (b), respectively. 

We will discuss first the trends of the core-electron eigenvalues of the
non-excited systems in their electronic ground states, i.e., the initial-state
contributions to the ACLSs, $\Delta^{\prime \rm initial}_{\rm ACLS}$.
The calculations show that the initial-state core-electron binding
energies of Cu atoms in the  
Pd$_{1-x}$Cu$_{x}$ overlayer on Cu are lower  
than those of the surface atoms of the clean Cu surface,
i.e., a negative
initial-state shift
(see Fig. 1(a)).
The magnitude of the shift increases with 
increasing foreign atom (Pd) concentration.
Similar results are obtained  for the
Ag core electrons in
Pd$_{1-x}$Ag$_{x}$  on Ag(001) relative to the clean Ag(001) surface.
Also for the $3d$ core electrons of Pd in Cu$_{1-x}$Pd$_{x}$ on Pd
(see Fig. 1(b))
and in
Ag$_{1-x}$Pd$_{x}$ on Pd,
we find that the initial-state shift relative to clean Pd surface
is negative.
Inspection of the surface $d$ DOS of the considered B-atoms
in the random
A$_{1-x}$B$_{x}$
overlayers shows for all four systems
that the $d$ band width decreases with increasing concentration of foreign
atoms, being narrowest for the single-impurity limit $(x\rightarrow 0)$. 
This result is indeed plausible and reflects the fact that coupling of $d$
orbitals between atoms is best when the neighbors are alike.
Thus, the more distinct the neighbors are, the worse is the coupling and the
stronger is the localization of the $d$ states.
Fig. 2(a) displays the $d$ contribution of the surface DOS
of Cu atoms in the random
Pd$_{1-x}$Cu$_{x}$ overlayer on Cu(001) for selected
concentrations and Fig. 2(b) the corresponding result for Pd atoms in
Cu$_{1-x}$Pd$_{x}$ on Pd(001). 
It is by now well established the essential
correctness of the band-narrowing
argument \cite{citrin},
i.e., that a narrowing of the $d$ band of a surface
atom is accompanied by a $d$ band shift in order to preserve $d$ band 
filling. The mentioned $d$ band-narrowing and shift to lower binding energies
as the concentration of foreign atoms increases is clearly visible in
Figs. 2(a) and (b).
It should be noted that similarly to the case of clean
surfaces \cite{clean} we have found a correlation between the
initial-state contributions to the ACLSs and the shift of the
$d$ band center as measured by the corresponding shift of the 
LMTO potential parameter $C_d$, athough deviations increase 
with the concentration of foreign atoms in the overlayer.

When final-state effects are taken into account, i.e.
differences in the
screening of the core hole by the other electrons,
we find that  the simple initial-state picture is changed considerably.
Although it is often argued that such screening contributions
are not very important (see e.g. Refs. \cite{weinert,initial})
our explicit calculations clearly identify here the contrary.
It has been already stressed 
that for clean surfaces and for pure
metallic adlayers the screening contribution can amount to some
tenths of an eV \cite{henn96,skriver94,meth95,gay82}.
Here, for the surface alloy case, we find
that the contributions due to differences in screening of the core-hole
due to modifications of the chemical environment of the adatoms,
can even change the qualitative trend of the initial-state approximation.
The striking example is shown in Fig. 1(b). 

Using the system Cu$_{1-x}$Pd$_{x}$/Pd(001) as an example, the 
screening contributions to the ACLSs of a Pd atom in the random 
Cu$_{1-x}$Pd$_{x}$ overlayer on Pd(001) are given by

\begin{eqnarray}
\Delta_{ACLS}^\prime -\Delta_{ACLS}^{\prime\, \rm initial}& = &
\Delta\epsilon_{c}^{{\rm Cu}_{1-x}{\rm Pd}_{x}/{\rm Pd(001)}}
 -\Delta\epsilon_{c}^{\rm Pd(001)} \quad , \label{eq3}
\end{eqnarray}

\noindent where $\Delta\epsilon$ is the positive energy by which the 
core eigenvalue drops when half an electron  
is removed. A large positive screening contribution to the 
Pd $3d$ ACLSs (see Fig. 1(b)) means 
that the Pd core levels in the random overlayer drop more strongly
than at the clean surface when deoccupied.
Thus, the Pd $3d$ core-hole is less well screened in  the random 
Cu$_{1-x}$Pd$_{x}$
overlayer.
A similar situation is found for Pd in    
Ag$_{1-x}$Pd$_{x}$ on Pd(001), i.e., 
the screening at the clean Pd surface is better than in the random 
Ag$_{1-x}$Pd$_{x}$ 
overlayer.
On the other hand, for Cu in the random
Pd$_{1-x}$Cu$_{x}$ overlayer on Cu(001) 
a negative screening
contribution is obtained meaning that the Cu $2p$ levels are
better screened in the random overlayer (see Fig. 1(a)). For Ag in 
Pd$_{1-x}$Ag$_{x}$ on Ag(001) 
a very small but positive contribution is obtained.

These results can be understood in terms of the explanations
developed in Ref.\cite{meth95} where it was noted that
screening at surfaces of true (more than one $d$ hole)
transition metals is better than in the bulk,
whereas at surfaces of noble metals it is worse. Generalizing 
their analysis to the more complex situation of a surface alloy we
conclude the following.
The noticeably better 
screening
of the Pd $3d$ core-hole
at the clean Pd surface than  in
the random Cu$_{1-x}$Pd$_{x}$ and Ag$_{1-x}$Pd$_{x}$ overlayers
is partially due to a
worsening
of the intra-atomic
screening
upon alloying, i.e., the screening  provided by the localized $d$ states on 
the atom with the core-hole. In fact, 
the $d$ DOS of the Pd atoms at the Fermi level
decreases up to 60\% ($x\rightarrow 0$) at the random overlayers.
In addition the interatomic screening, i.e., screening 
provided by surrounding atoms,   
also becomes
worse upon alloying with noble metals such as Cu or Ag.
On the other hand, for  noble metals such as Cu or Ag 
intra-atomic 
screening is less efficient due to the lower DOS at the
Fermi level. 
Thus, interatomic screening becomes important, which
is reflected by the 
better screening of the Cu $2p$ hole in the random 
Pd$_{1-x}$Cu$_{x}$ overlayer as the Pd concentration increases. 
The final-state screening contributions for Ag in Pd$_{1-x}$Ag$_{x}$ on Ag
are very small but positive.
Though one could have expected  both Cu and Ag core holes to be     
equally screened when surrounded by Pd atoms,
one should notice
 that Pd atoms in a Cu matrix as in Pd$_{1-x}$Cu$_{x}$/Cu(001) are 
considerably more compressed than 
in an Ag matrix (where interatomic distances are actually 5\% 
larger that at a clean Pd surface). 
Thus, interatomic screening is facilitated in the first case.
Inspection of the Pd $d$ DOS 
of the random overlayers clearly shows the effect
of the different hosts on the Pd $d$ band, being  much wider for
Pd in Pd$_{1-x}$Cu$_{x}$ on Cu(001) than for the clean Pd(001) surface, which
in turn is wider than for Pd  in Pd$_{1-x}$Ag$_{x}$ on Ag(001).

In summary, the results presented here unambiguously show the 
importance of final-state screening effects for understanding and 
interpreting measured values of core-level shifts. 
This conclusion agrees with some previous studies
\cite{henn96,skriver94,meth95,pehl93,gay82} for ideal surfaces or
admetal monolayers.
In the present case, screening contributions originate from differences in the 
screening of the core-hole due to modifications of the chemical 
environment of the overlayer atoms.
The present calculations, which systematically separate 
initial- and final-state effects, show 
that the measured core-level shifts do in fact 
not necessarily correlate with the initial-state shifts and therefore
they do not necessarily reflect the shift of the
surface $d$ DOS. The results support and stress
the conclusion that the shift of the surface $d$ band is
just an initial-state effect  and the use of measured values
of ACLS's as an indication of the surface $d$ band shift is, in
general, incorrect. 
We have also performed similar calculations for the Cu (Ni) $2p$ 
levels in the CuNi random overlayers on Cu (Ni). The difference in the
efficiency of the screening between transition and noble metals is confirmed.
Thus, the Cu (Ni) $2p$ core-hole is better (worse) screened in the
random CuNi overlayers than at the clean Cu (Ni) surface.
The increased  (negative) initial-state contributions to the
ACLSs of Pd in the random overlayer of 
Cu$_{1-x}$Pd$_{x}$ on Pd(001) (see Fig. 1(b))
and in Ag$_{1-x}$Pd$_{x}$ on Pd(001) as the 
concentration of Pd decreases, which are accompanied by a higher 
Pd $d$ DOS closer to the Fermi level (see Fig. 2(b)), suggest
that alloying an active metal like Pd with a
catalytically inactive component, such as Cu or Ag, 
would result in an 
{\em increased} chemical reactivity. 
Our predictions relate only to those changes
in activity that can be interpreted in terms
of an  electronic effect and our above analysis
would
provide basic information on the influence of alloying on the surface
reactivity only if the electronic effect is the dominant effect.
It would be interesting to see  if these predictions will be verified
experimentally.
Finally, we note that
recently
the variations in the surface electronic structure
for  supported monolayers as in Ref. \cite{henn96}
and substitutional impurities for a whole set of systems have
been  studied and similar predictions have
been made \cite{ruban}.

M.V. Ganduglia-Pirovano thanks J. Norskov, A. Ruban and M.H. Cohen for
useful discussions.  
One of us (J.K.) acknowledges the financial support from the
US-Czechoslovak Science and Technology program (project No. 95018).
Center for Atomic-scale Materials Physics is sponsored by the Danish National
Research Foundation.

\begin{figure}
  \leavevmode
  \includegraphics{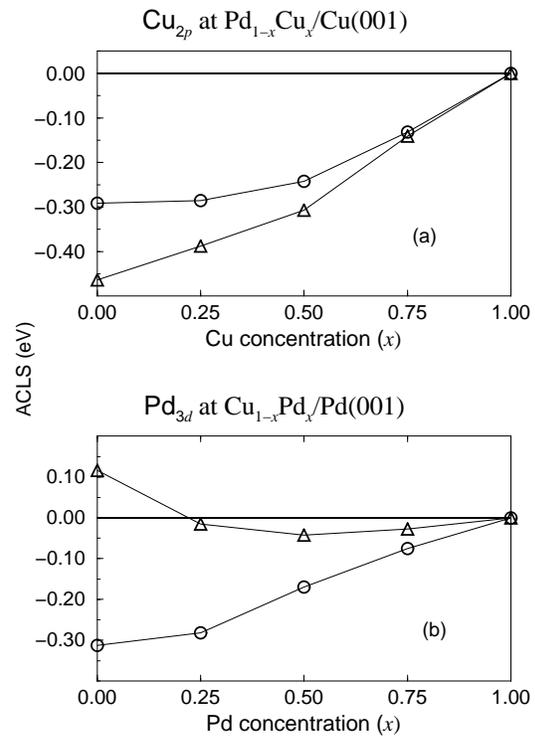}
  \vspace*{7.5cm}
\caption{(a) The ACLSs of Cu $2p$ electrons in the random 
Pd$_{1-x}$Cu$_{x}$ overlayer on Cu(001) (triangles)
(see Eq. \protect{\ref{eq2}})
and the initial-state contributions to the shifts (circles).
The limit of zero concentration means a
single Cu impurity in a Pd monolayer on Cu(001). (b) 
is similar to (a) but for the Pd $3d$ electrons  in
Cu$_{1-x}$Pd$_{x}$ on Pd(001). 
}\label{fig.1}
\end{figure}

\begin{figure}
  \leavevmode
  \includegraphics{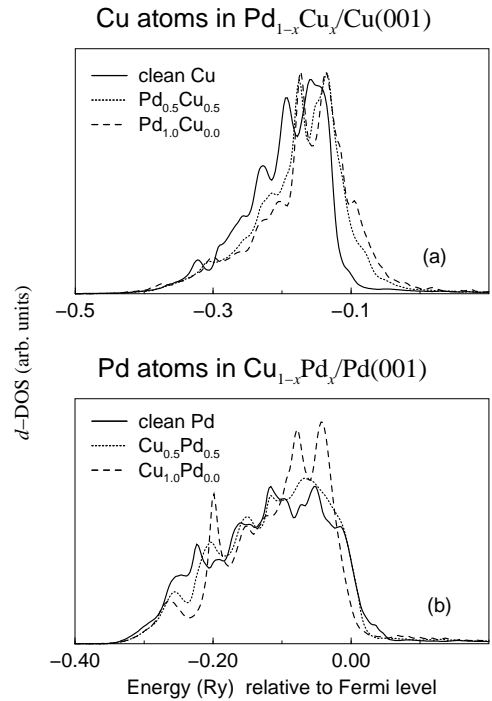}
  \vspace*{7.5cm}
\caption{(a) The $d$-DOS for Cu atoms in the random 
Pd$_{1-x}$Cu$_{x}$ overlayer on Cu(001) for selected 
concentrations.
The limit of zero concentration means a
single Cu impurity in a Pd monolayer on Cu(001). (b) 
is similar to (a) but for the Pd  atoms in   
Cu$_{1-x}$Pd$_{x}$ on Pd(001). 
}\label{fig.2}
\end{figure}


\vspace*{-0.5cm}

\begin{references}
\bibitem{sinfelt}
   J. H. Sinfelt, {\it Bimetallic Catalysts} (Wiley, N.Y., 1983).
\bibitem{goodman1}
   J. A. Rodriguez, R. A. Campbell, and D. W. Goodman,
   J. Vac. Sci. Technol. A {\bf 10}, 2540 (1992);
   Surf. Sci. {\bf 307-309}, 377 (1994).
\bibitem{weinert}
M. Weinert, R.E. Watson, Phys. Rev. B {\bf 51}, 17168 (1995).
\bibitem{wu}
R. Wu, A.J. Freeman, Phys. Rev. B {\bf 52}, 12419 (1995).
\bibitem{henn96}
D. Hennig, M.V. Ganduglia Pirovano, and M. Scheffler, Phys. Rev.   
B {\bf 53}, 10344 (1996).
\bibitem{hamm95a}
    B. Hammer and M. Scheffler, Phys. Rev. Lett. {\bf 74}, 3487 (1995).
\bibitem{hammer95}
    B. Hammer and J.K. Norskov, Surf. Sci. {\bf 343}, 211 (1995);
    B. Hammer, Y. Morikawa, and J.K. Norskov, Phys. Rev. Lett. {\bf 76}, 2141 (1996).
\bibitem{skriver94}
M. Alden, H.L. Skriver and B. Johansson, Phys. Rev. Lett. {\bf 71}, 2449
(1993);
M. Alden, I.A. Abrikosov, B. Johansson, N.M. Rosengaard, and H.L. Skriver,
 Phys. Rev. B {\bf 50}, 5131 (1994).
\bibitem{meth95}
   M. Methfessel, D. Hennig, and M. Scheffler, Surface Reviews and Letters
   {\bf 2}, 197 (1995).
\bibitem{pehl93} E. Pehlke, M. Scheffler,
Phys. Rev. Lett. {\bf 71}, 2338 (1993).
\bibitem{ts}
J. C. Slater, in {\it Quantum Theory of Molecules and Solids},
(Mc Graw-Hill, New York, 1974), vol. 4, pp. 51-55; J.F. Janak, Phys. Rev. 
B {\bf 18}, 7165 (1978).
\bibitem{josef1}
J.  Kudrnovsk\'{y}, I. Turek, V. Drchal, P. Weinberger, N.E. Christensen,
and S.K. Bose, Phys. Rev. B {\bf 46}, 4222 (1992).
\bibitem{clean}
    M.V. Ganduglia-Pirovano, V. Natoli, M.H. Cohen, J. Kudrnovsk\'{y}, and I. 
Turek, Phys. Rev. B. {\bf 54}, 8892 (1996).
\bibitem{citrin}
    P.H. Citrin, G.K. Wertheim and Y. Baer, Phys. Rev. Lett. {\bf 41},
    1425 (1978); M.C. Desjonqueres {\it et al.}, Sol. State Comm. {\bf 34}, 807
    (1980); P.H. Citrin and G.K. Wertheim, Phys. Rev. B {\bf 27}, 3176 (1983).
\bibitem{meth93}
   M. Methfessel, D. Hennig, and M. Scheffler, Surf. Sci. {\bf 287/288}, 785 
(1993).
\bibitem{lmto}
    H. L. Skriver, {\it The LMTO Method}, (Springer-Verlag, Berlin, 1984).
\bibitem{initial}
P.J. Feibelman and D.R. Hamman,
Sol. State Commun. {\bf 31}, 413 (1979);
Phys. Rev. B {\bf 28}, 3092 (1983); P.J. Feibelman, Phys.
Rev. B {\bf 39}, 4866 (1989).
\bibitem{gay82}
J.R. Smith, F.J. Arlinghaus, and J.G. Gay, Phys. Rev. B {\bf 26}, 1071 (1982).
\bibitem{ruban}
A. Ruban, B. Hammer, P. Stoltze, H.L. Skriver, and J.K. Norskov,
J. Mol. Catalysis A: Chemical {\bf 115}, 421 (1997).
\end{references}
\end{document}